\documentclass[twocolumn,showpacs,preprintnumbers,amsmath,amssymb,pra]{revtex4-1}
\usepackage{amsfonts}
\usepackage{mathrsfs}
\usepackage{graphicx}
\usepackage{dcolumn}
\usepackage{bm}

\begin{document}
\title{Correlation Properties of anisotropic XY Model with a Sudden Quench}
\author{Hongli Guo}
\address{Institute of Physics, Chinese Academy of Sciences, Beijing 100190, China}
\author{Zhao Liu}
\address{Institute of Physics, Chinese Academy of Sciences,
Beijing 100190, China}
\author{Heng Fan}
\address{Institute of Physics, Chinese Academy of Sciences,
Beijing 100190, China}
\author{Shu Chen}
\address{Institute of Physics, Chinese Academy of Sciences,
Beijing 100190, China}
\date{\today}

\begin{abstract}
Starting from a general Hamiltonian which may undergo a quantum
phase transition (QPT) with the change of a controllable parameter,
we obtain a general conclusion that in a sudden quench system, when
the final Hamiltonian is fixed, the behavior of the time-averaged
expectation of any observable has close relationship with the
gapless excitation of the initial Hamiltonian. To clarify our
conclusion, we investigate the two-spin correlation of a XY chain in
a transverse field under a sudden quench at zero temperature. The
critical property of the derivative of quench two-spin correlation
and the long-range correlation of the quench system are analyzed.
\end{abstract}

\maketitle


\section{Introduction}
Quantum phase transition (QPT), which is driven by quantum
fluctuation and occurs at zero temperature, is a very important
research area recently \cite{book}. Besides the static properties of
QPT, the non-equilibrium dynamics induced by the quench of the
parameter in the Hamiltonian through a critical point has always
been an attractive topic in condensed matter physics
\cite{Na41951,PRA69053616,PRL95105701,PRL95245701,PRB72161201,PRA72052319,PRA73063405,PRB76174303,PRA75023603,PRL100077204,Np4
477,nature452,PRB80054404,PRL98050405,PRL102245701,nature449324,PRA79021608}.

Generally speaking, a dynamical evolution can be induced by either a sudden quench or
a slow quench. Taking the spin chain in the magnetic field as
an example, the typical paradigm of sudden quench is as follows. Initially ($t<0$) the magnetic
field $h(t<0)=a$. Then at time $t=0$ the magnetic field is changed to $b$ suddenly, namely $h(t\geq0)=b$
and the system begins to evolve from the initial state. After a long enough time evolution, the
time-averaged expectation value of an observable $A$ reaches a
steady value, which we define as quench $A$. Quench $A$ is $\langle A(a,
b)\rangle=\lim_{t\rightarrow\infty}\frac{1}{t}\int_{0}^{t}\langle
A(a, b, \tau)\rangle d\tau$ \cite{nature452}, with $\langle
A(a,b,\tau)\rangle$ the expectation value of observable $A$ at time
$t=\tau$ with initial magnetic field $a$ and final magnetic field
$b$. One can fix $b$($a$) to investigate the relation between quench $A$ and $a$($b$).

The magnetization $M$ and its derivative with respect to magnetic
field $\partial_{h}M$ of transverse Ising model are analyzed in Ref.
\cite{PRB80054404}. It is found that $\partial_{h}M$ has similar
behavior for static system and sudden quench system. Here
interesting questions arise: why does this similarity happen and
whether other quantities still exhibit such similarity?

In this work, we start from a general Hamiltonian $H(\lambda)$
containing a controllable parameter $\lambda$ to search for the
answers. We obtain a general conclusion that in a sudden quench
system, when the final Hamiltonian is fixed, the behavior of the
time-averaged expectation of any observable has close relationship
with the gapless excitation of the initial Hamiltonian. There is a
similarity between the critical phenomena of sudden quench and
static system. Then we numerically investigate the sudden quench
properties of two-spin correlation of an anisotropic XY chain in a
transverse field \cite{PRA21075}. The results are consistent with
our conclusion. We also research the long-range correlation of the
system in both directions parallel and perpendicular to the magnetic
field. There is no qualitative difference between static and sudden
quench case in the direction parallel to the magnetic field, but a
remarkable difference exists in the direction perpendicular to the
magnetic field, which is consistent with the result in
Ref.\cite{PRA69053616}. At last, we briefly discuss the case in
which initial Hamiltonian is fixed.

\section{General Model}
Here we consider a general Hamiltonian $H(\lambda)=H_{0}+\lambda
H_{I}$ which contains a controllable parameter $\lambda$. We suppose
the ground state of $H(\lambda)$ is non-degenerate and with the
varying of $\lambda$ the system can undergo a QPT happening at
$\lambda=\lambda_{c}$, where the energy gap of $H(\lambda)$ vanishes in the
thermodynamics limit (TL). This is the case for many condensed
matter systems, especially for spin chains. For example
$H_{I}=\sum_{i}\sigma_{i}^{z}$ and $\lambda$ is the magnetic field
for transverse Ising model.

Now we try to derive the singular behavior of the time-averaged
expectation value of any observable $A$ when QPT happens, both for
static case and sudden quench case. At time $t<0$, $\lambda=a$ and
the eigenvectors and eigenvalues of $H(a)$ are denoted as
$|\psi_{n}(a)\rangle$ and $E_{n}(a)$ respectively. At zero
temperature the system is in the ground state of this initial
Hamiltonian $H(a)$, namely $|\Psi(t<0)\rangle=|\psi_{0}(a)\rangle$.
Therefore for static case, which means $\lambda=a$ for $t\geq0$, the
the time-averaged expectation value is just $\langle
A(a)\rangle=\langle\psi_{0}(a)|A|\psi_{0}(a)\rangle$. By the
perturbation theory, we know
\begin{eqnarray}
|\psi_{0}(a+\delta a)\rangle=|\psi_{0}(a)\rangle+\delta a\sum_{m\neq
0}\frac{(H_{I})_{m0}}{E_{0}(a)-E_{m}(a)}|\psi_{m}(a)\rangle\nonumber
\end{eqnarray}
with $(H_{I})_{m0}=\langle \psi_{m}(a)|H_{I}|\psi_{0}(a)\rangle$, so
\begin{eqnarray}
\partial_{a}|\psi_{0}(a)\rangle=\sum_{m\neq
0}\frac{(H_{I})_{m0}}{E_{0}(a)-E_{m}(a)}|\psi_{m}(a)\rangle,\nonumber
\end{eqnarray} which
leads to
\begin{eqnarray}
\partial_{a}\langle
A(a)\rangle=2\sum_{m\neq0}\frac{\Re[(H_{I})_{0m}\langle
\psi_{m}(a)|A|\psi_{0}(a)\rangle]}{E_{0}(a)-E_{m}(a)}\label{e4}
\end{eqnarray}
straightforwardly, where $\Re(z)$ means the real part of a complex
number $z$. One can expect that $\partial_{a}\langle A(a)\rangle$
will behave singularly at $a=\lambda_{c}$ because the energy gap
vanishes in TL.

If at time $t=0$, $\lambda$ is suddenly changed to $b$ and
$\lambda=b$ is kept for $t>0$, the system will evolve under sudden
quench. The eigenvectors and eigenvalues of the final Hamiltonian
$H(b)$ are denoted as $|\phi_{n}(b)\rangle$ and $\omega_{n}$
respectively. The time-evolving wave function of the system at $t>0$
is $|\Psi(t>0)\rangle=\sum_{n}\langle \phi_{n}(b)|\psi_{0}(a)\rangle
e^{-i \omega_{n}t}|\phi_{n}(b)\rangle$, where we omit the relative
phase. The time-averaged expectation value is
\begin{eqnarray}
\langle
A(a,b)\rangle&=&\lim_{t\rightarrow\infty}\frac{1}{t}\int_{0}^{t}\langle
A(a,b,\tau)\rangle d\tau\nonumber\\
&=&\sum_{n}\langle \psi_{0}(a)|\phi_{n}(b)\rangle\langle
\phi_{n}(b)|\psi_{0}(a)\rangle\langle
\phi_{n}(b)|A|\phi_{n}(b)\rangle.\nonumber
\end{eqnarray}
So we have
\begin{widetext}
\begin{eqnarray}
\partial_{a}\langle A(a,b)\rangle
&=&2\sum_{m\neq0}\sum_{n}\frac{\langle
\phi_{n}(b)|A|\phi_{n}(b)\rangle}{E_{0}(a)-E_{m}(a)}\times\Re[(H_{I})_{0m}\langle
\psi_{m}(a)|\phi_{n}(b)\rangle \langle
\phi_{n}(b)|\psi_{0}(a)\rangle]. \label{e5}
\end{eqnarray}
\end{widetext}
 When $b\neq \lambda_{c}$, the
numerator in the above Eq.(\ref{e5}) is not zero. Therefore
$\partial_{a}\langle A(a,b)\rangle$ will show singularity at
$a=\lambda_{c}$ because of the vanishing of the energy gap in TL.
However if $b=\lambda_{c}=a$, we have $\psi=\phi$, so in
Eq.(\ref{e5}) $\langle \psi_{m}(a)|\phi_{n}(b)\rangle \langle
\phi_{n}(b)|\psi_{0}(a)\rangle=\delta_{m,n}\delta_{n,0}$.
Considering $m\neq 0$ in the sum, we can see both the numerator and
denominator in Eq.(\ref{e5}) are zero at $a=b=\lambda_{c}$, meaning
that $\partial_{a}\langle A(a,\lambda_{c})\rangle$ maybe has no
singular behavior at $a=\lambda_{c}$.

Because Eqs.(\ref{e4}) and (\ref{e5}) share the same denominator, we
conclude that there must be similarity between the critical singular
behaviors of $\partial_{a}\langle A(a)\rangle$ and
$\partial_{a}\langle A(a,b)\rangle$. The singularity and scaling
behaviors come from the vanishing of the energy gap of the initial
Hamiltonian in TL. If there is a critical phenomenon of the initial
Hamiltonian with a vanishing energy gap which can be indicated by
the singular behavior of $\partial_{a}\langle A(a)\rangle$ at
$a=\lambda_{c}$, $\partial_{a}\langle A(a,b)\rangle$ will also have
a similar singular behavior at $a=\lambda_{c}$ if
$b\neq\lambda_{c}$.

\section{XY spin chain model}
In this paper, we use the anisotropic XY model in a transverse field
under a sudden quench at zero temperature to check our conclusion in
the above section. The Hamiltonian of the spin-1/2 anisotropic XY
spin chain is
\begin{eqnarray}
H=\sum_{j=0}^{N-1}\Big\{\frac{J}{2}\Big[(1+\gamma
)\sigma_{j}^{x}\sigma_{j+1}^{x}+(1-\gamma
)\sigma_{j}^{y}\sigma_{j+1}^{y}\Big]-h(t)\sigma_{j}^{z}\Big\},\nonumber
\label{1}
\end{eqnarray}
where $J$ is the coupling constant,
$\sigma_{j}^{x},\sigma_{j}^{y},\sigma_{j}^{z}$ are the Pauli
operators at the $j$th lattice site and $J(1\pm\gamma)/2$ is the
measure of the interaction strength in $x(y)$ component between two
nearest-neighbor spins. $\gamma$ is the anisotropy parameter which
can change from $0$ (XX model) to $1$ (Ising model). $h(t)$ is the
time-dependent magnetic field. In our paper, we set $J=1>0$ for
convenience and the periodic boundary condition $\sigma_{N}^{\alpha
}=\sigma_{0}^{\alpha }, \alpha =x,y,z$ is used. The phase diagram of
this model is easy to be obtained. There are two kinds of phase
transition in anisotropic XY spin chain model. One belongs to the
universality class of the Ising phase transition and the other is
anisotropic transition \cite{PRB76174303}. In this paper we focus on
the Ising phase transition. The equilibrium Ising phase critical
points are $|h|=1$. In $J=1>0$ case, the system has
anti-ferromagnetic interaction with the staggered magnetization in
the $x$ direction
$\langle\sigma^x\rangle_{\textrm{stag}}=\frac{1}{N}\sum_{i=0}^{N-1}
(-1)^i\langle\sigma_i^x\rangle$ as the order parameter. For $|h|<1$,
the system is in an ordered phase with
$|\langle\sigma^x\rangle_{\textrm{stag}}|=1$, while for $|h|>1$, the
system is in a disordered phase with
$\langle\sigma^x\rangle_{\textrm{stag}}=0$. Now we turn to the
problem of dynamics under sudden quench. Using successive
Jordan-Wigner, Fourier, and Bogoliubov transformations, the exact
solutions of the evolved state of XY model under a sudden quench can
be obtained \cite{PRA21075}, by which the quench dynamical
properties of many observables such as magnetization and two-spin
correlation can be studied. In our work we focus on the quench
two-spin correlation, which is defined as $G^{\alpha}_{n}(a,b)=
\lim_{t\rightarrow\infty}\frac{1}{t}\int_{0}^{t}\langle
\sigma^{\alpha}_{0}(a, b, \tau)\sigma^{\alpha}_{n}(a, b,
\tau)\rangle d\tau$, where $\alpha=x,y,z$, $n$ is the distance
between two spins and $a$ ($b$) is the initial (final) magnetic
field.

\section{Quench two-spin correlation}
As the beginning we want to clarify our conclusion in Sec.II by
considering $\partial_{a} G^{z(x)}_{1}(a,b)$, namely the
nearest-neighbor two-spin correlation. Due to the difficulty in the
analytical calculation of them, we choose to numerically solve them.
We let $\gamma=1$ for convenience in this part and one should note
that all conclusions below are also valid for $\gamma\neq1$.

\begin{figure}
\centerline{\includegraphics[width=\linewidth]{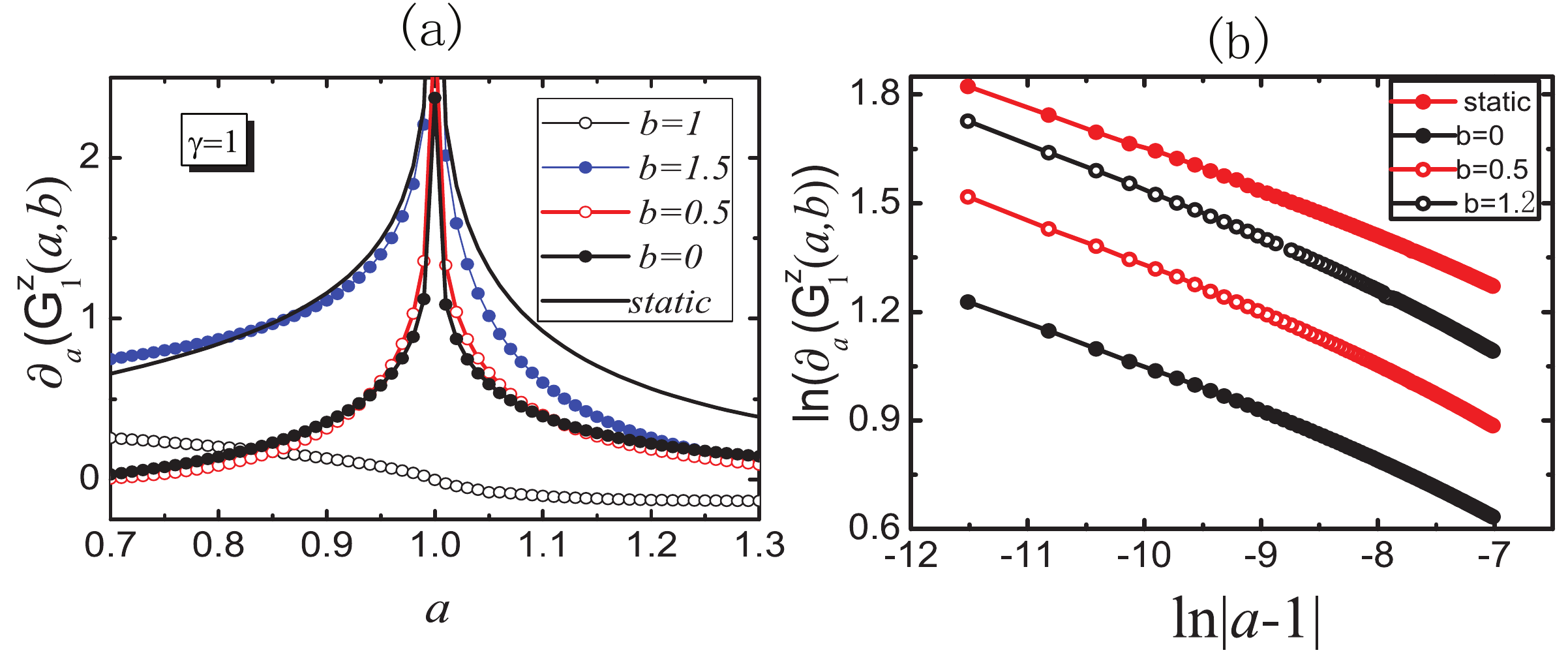}}
\caption{$\gamma=1$. (a) $\partial_{a} G^{z}_{1}(a,b)$ for static
case and dynamical cases $b=0, 0.5, 1, 1.5$. (b) $\ln(\partial_{a}
G^{z}_{1}(a,b))$ as a function of $\ln|a-1|$. One can see the slopes
of the lines for different $b$ are all the same, namely a universal
constant -0.25.}\label{f6}
\end{figure}

\begin{figure}
\centerline{\includegraphics[width=\linewidth]{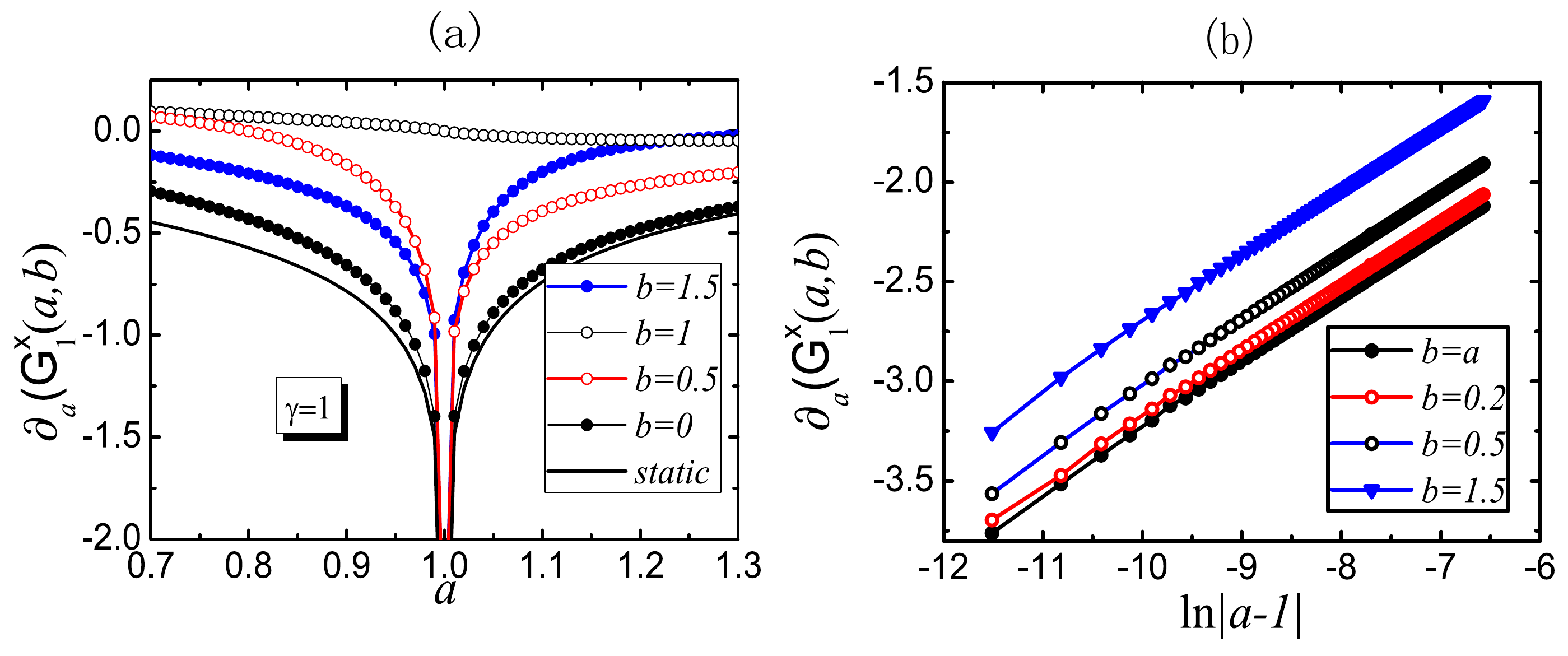}}
\caption{$\gamma=1$. (a) $\partial_{a} G^{x}_{1}(a,b)$
for static case and dynamical cases $b=0, 0.5, 1, 1.5$. (b)
$\partial_{a} G^{x}_{1}(a,b)$ as a function of $\ln|a-1|$. One can
see the slopes of the lines for different $b$ are all the same,
namely a universal constant $1/\pi$.}\label{f9}
\end{figure}

In Fig.\ref{f6}(a), $\partial_{a} G^{z}_{1}(a,b\neq1)$ diverges at
$a=1$ for both the static and dynamical case. Moreover, we calculate
that $\partial_{a} G^{z}_{1}(a,b)=\textrm{const}\times|a-1|^{-\mu}$
with $\mu\approx0.25$, a universal exponent irrelevant with $b$
(Fig.\ref{f6}(b)). While as predicted in Sec.II, the sudden quench
case with $b=1$ is very special. We can see from Fig.\ref{f6}(a)
that $\partial_{a} G^{z}_{1}(a,1)$ is not divergent at $a=1$.
Similar to $G_{1}^{z}(a,b)$, $\partial_{a}G^{x}_{1}(a,b\neq1)$
diverges at $a=1$ for static and sudden quench case. After numerical
analysis, we can find $\partial_{a}
G^{x}_{1}(a,b)=\frac{1}{\pi}\ln|a-1|+f(b)$, with $f(b)$ a constant
which we do not concern with (Fig.\ref{f9}(b)). We also can see from
Fig.\ref{f9}(a) that $\partial_{a} G^{x}_{1}(a,1)$ is not divergent
at $a=1$.

Through the comparison between static and sudden quench cases of XY
model, we find that in general sudden quench case ($b\neq1$), the
quench system and static system share similar critical scaling form
of $\partial_{a} G^{z(x)}_{1}(a,b)$. This is consistent with our
conclusion in Sec.II and reflects the memory of the quench system to
the initial Hamiltonian.

\begin{figure}
\centerline{\includegraphics[width=\linewidth]{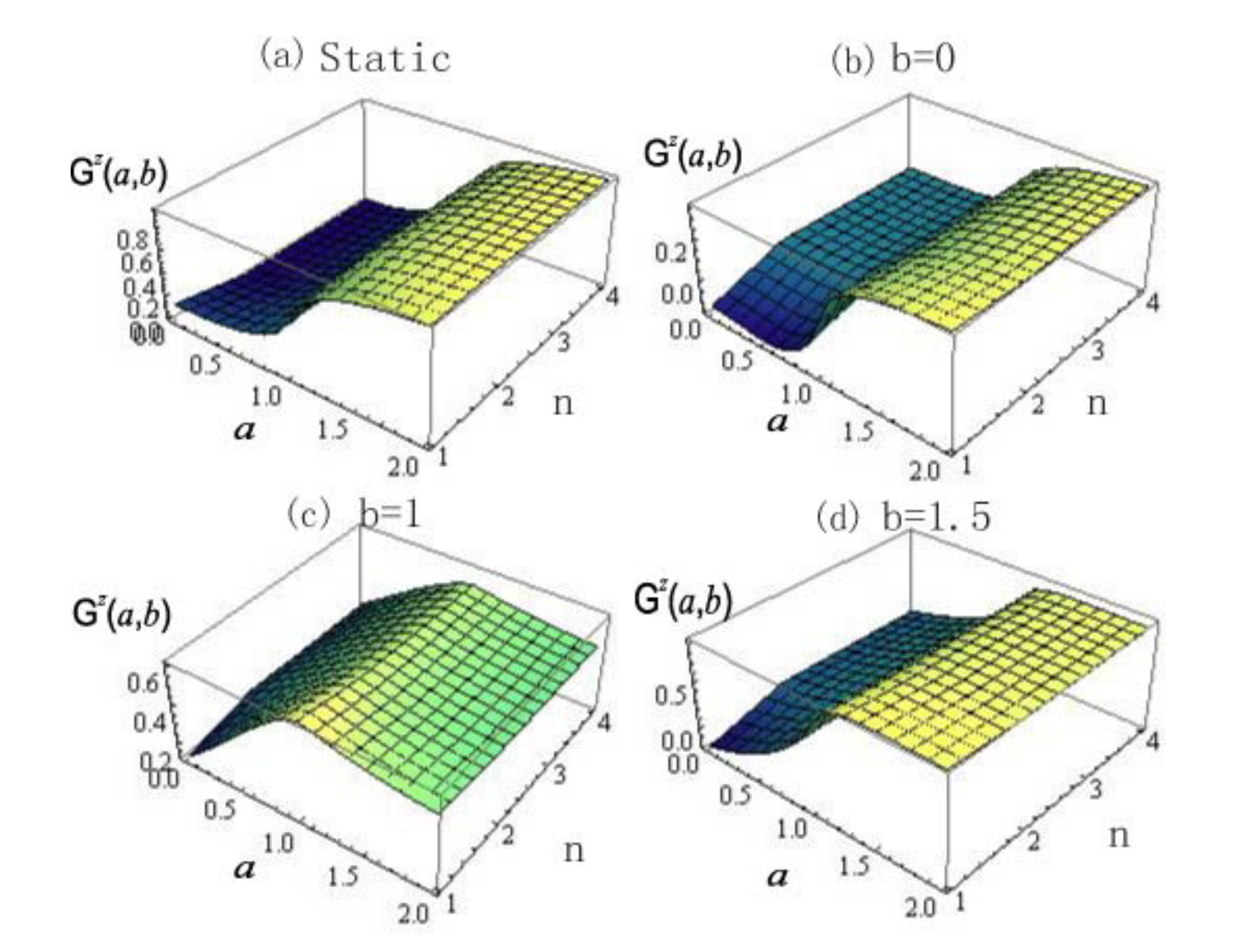}}
 \caption{$\gamma=0.6$. $G^{z}_{n}(a,b)$ for
(a) The static case $b=a$, (b) $b=0$, (c) $b=1$ and (d) $b=1.5$.}
\label{f5}
\end{figure}

Now we shift our focus from the nearest-neighbor two-spin
correlation to the correlation between two spins with arbitrary
distance $n$. We show $G^{z}_{n}(a,b)$ as a function of $a$ and $n$
for both static and sudden quench case in Fig.\ref{f5}, where
$\gamma=0.6$ is used. From this figure, we know as $n$ increases,
$G^{z}_{n}(a,b)$ reaches a constant quickly after a little increase
of $n$. There is no qualitative difference between static and
general dynamical case.

\begin{figure}
\centerline{\includegraphics[width=\linewidth]{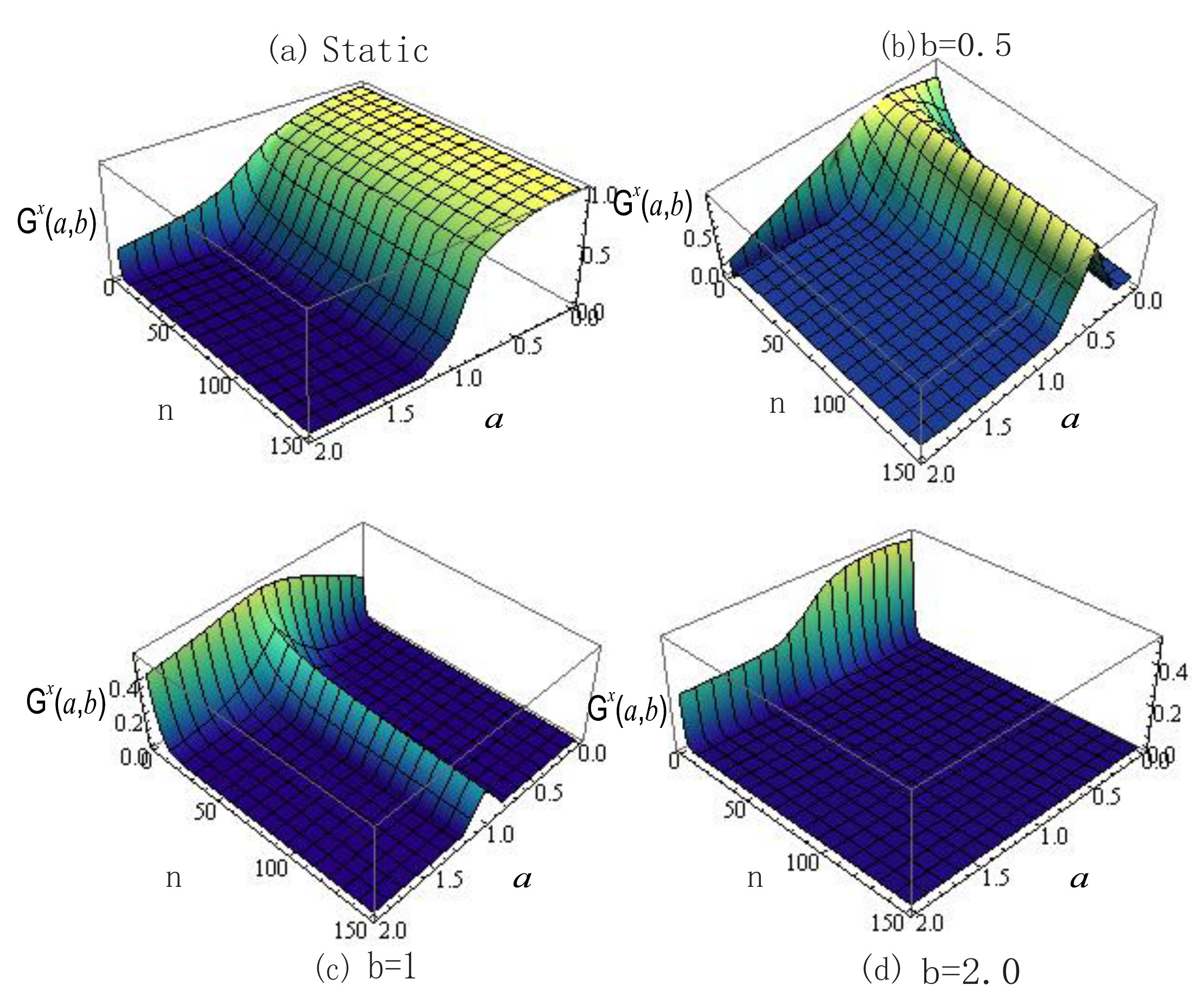}}
 \caption{$\gamma=1$. $G^{x}_{n}(a,b)$ in (a) static system and
dynamical systems with (b) $b=0.5$, (c) $b=1$ and (d)
$b=2$.}\label{f7}
\end{figure}

\begin{figure}
\centerline{\includegraphics[width=\linewidth]{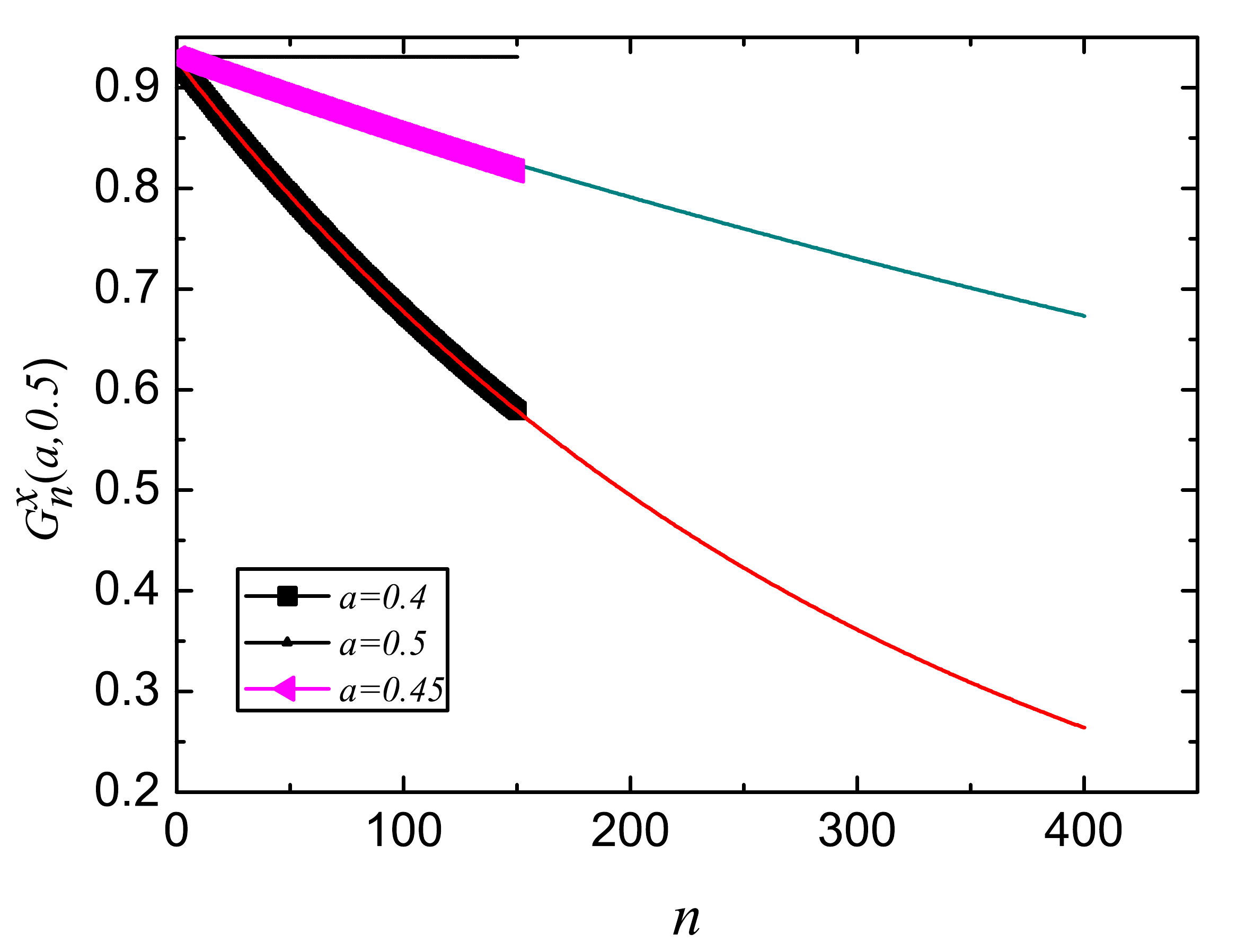}}
\caption{Numerical fitting of $G_n^x(a,b)$ at $b=0.5$ up to $n=150$.
Through numerical fit of the two typical data $G^{x}_{n}(0.4,0.5)$
and $G^{x}_{n}(0.45,0.5)$, we find $G_n^x(a,b)\sim
n^{-c_{1}}e^{-c_{2}n}$, with
$G_n^x(0.4,0.5)=n^{-0.0000404236}e^{(-0.074916n-0.0031414)}$ and
$G_n^x(0.45,0.5)=n^{-0.0000460715}e^{(-0.07259538n-0.00080686)}$. We
extrapolate those two curves to $n=400$.}\label{f8}
\end{figure}

For $G^{x}_{n}(a,b)$, there will be a big difference. We let
$\gamma=1$ for convenience here. In a static system ($a=b$), from
Fig.\ref{f7}(a) we can see when $a>1$, $G^{x}_{n}(a)$ tends to zero
very quickly and is nonzero only if $n$ is very small, demonstrating
no long-range correlation in $x$ direction. While in another phase
$a<1$, the ground state is degenerate and possesses long-range
correlation in $x$ direction. However, in a sudden quench system,
the case will be obviously different. If the final magnetic field
$b<1$, after a long time evolution the long-range correlation in $x$
direction only exists at the point $a=b<1$ (that is just the static
case), meaning that when $b<1$ a change of magnetic field at $t=0$
will eliminate the long-range correlation in $x$ direction. In
Fig.\ref{f7}(b), we plot the $G^{x}_{n}(a,b)$ up to $n=150$ with
$b=0.5$. Through numerical fit of the two typical data
$G^{x}_{n}(0.4,0.5)$ and $G^{x}_{n}(0.45,0.5)$ (Fig.\ref{f8}), we
find $G_n^x(a,b)\sim n^{-c_{1}}e^{-c_{2}n}$, decaying to zero with
$n$ even when $a$ and $b$ is slightly different. So we think the
$G_n^x(a,b)$ is destroyed after an arbitrarily small quench. While
$b>1$, no matter what $a$ is, even for $a<1$, there is no long-range
correlation in $x$ direction. This means that there is a possibility
that a system possessing a long-range correlation in $x$ direction
at the beginning will lose it after a long enough time evolution
(Fig.\ref{f7}(d)). So, only when static case $a=b<1$ , the
long-range correlation in $x$ direction is preserved. In other cases
we can have when $n\rightarrow\infty$, $\langle\sigma_{0}^{x}(a,
b)\sigma_{n}^{x}(a,b)\rangle\rightarrow0$, leading to
$\langle\sigma_{x}(a,b)\rangle=0$ straightforwardly. A physical
intuition tells us that because the magnetic field is in the $z$
direction, the long-range correlation in the $z$ direction can
survive after the quench but the long-range correlation in the $x$
direction is destroyed by the quench.

We note that the long range correlation is also researched by
Sengupta \emph{et al}. in Ref.\cite{PRA69053616}. They give some
analytic results of the two point correlation function perpendicular
to the magnetic field direction in quantum Ising model. However,
they only consider two limit cases, namely the initial magnetic
field is fixed at $a=0$ and $a=\infty$. In both cases, $G_n^x(a,b)$
tends to zero when $n\rightarrow\infty$ but in different ways which
depend on the value of the final magnetic field $b$. In
Fig.\ref{f7}, one can see that $G_n^x(0,0.5)$ tends to zero much
slower than $G_n^x(0,2)$ and there is a clear spatial oscillation of
$G_n^x(2,0.5)$. These phenomena are consistent with
Ref.\cite{PRA69053616}. Therefore, our numerical results confirm
their analytic results about long range correlation and give an
extension to the general situation beyond the two limit cases.

\section{a further discussion: fixed initial magnetic field} Up to now we only
consider the case in which the final magnetic field $b$ is fixed and
all quench quantities are regarded as functions of initial magnetic
field $a$. For completeness, we discuss the case in which $a$ is
fixed and all quench quantities are regarded as functions of $b$. As
one can expect, the quench quantities will have different behaviors.
If we make a similar calculation with that in Sec.II using
perturbation theory, we can obtain
\begin{widetext}
\begin{eqnarray}
\partial_{b}\langle A(a,b)\rangle&=&2\sum_{n}\sum_{m\neq n}\frac{\langle
\phi_{n}(b)|A|\phi_{n}(b)\rangle}{\omega_{n}(b)-\omega_{m}(b)}\times\Re[(H_{I})_{nm}\langle
\psi_{0}(a)|\phi_{n}(b)\rangle\langle
\phi_{m}(b)|\psi_{0}(a)\rangle]\nonumber\\
&+&2\sum_{n}\sum_{m\neq n}\frac{\langle
\psi_{0}(a)|\phi_{n}(b)\rangle\langle
\phi_{n}(b)|\psi_{0}(a)\rangle}{\omega_{n}(b)-\omega_{m}(b)}\times\Re[(H_{I})_{nm}\langle
\phi_{m}(b)|A|\phi_{n}(b)\rangle]. \label{ee}
\end{eqnarray}
\end{widetext}
It is difficult to study the behavior of Eq.(\ref{ee}) near
$b=\lambda_{c}$ for a general $a$. However, the case in which
$a=\lambda_{c}$ is easy to analyze. When $a=\lambda_c=b$,
$\phi=\psi$. Therefore in the first term of the right hand side of
Eq.(\ref{ee}), $\langle \psi_{0}(a)|\phi_{n}(b)\rangle\langle
\phi_{m}(b)|\psi_{0}(a)\rangle=\delta_{m,0}\delta_{n,0}$, meaning
that this term is 0 because $m\neq n$ in the sum. Similarly, we have
$\langle \psi_{0}(a)|\phi_{n}(b)\rangle\langle
\phi_{n}(b)|\psi_{0}(a)\rangle=\delta_{n,0}$ in the second term of
Eq.(\ref{ee}). So we have
\begin{eqnarray}
\partial_{b}\langle
A(\lambda_{c},b)\rangle|_{b=\lambda_c}=2\sum_{m\neq
0}\frac{\Re[(H_{I})_{0m}\langle
\phi_{m}(\lambda_c)|A|\phi_{0}(\lambda_c)\rangle]}{\omega_{0}(\lambda_c)-\omega_{m}(\lambda_c)}.\nonumber
\end{eqnarray}
This equation is the same with Eq.(\ref{e4}) and means that
$\partial_{b}\langle A(\lambda_{c},b)\rangle$ will diverge at
$b=\lambda_c$ because the vanish of the energy gap.

\begin{figure}
\centerline{\includegraphics[width=\linewidth]{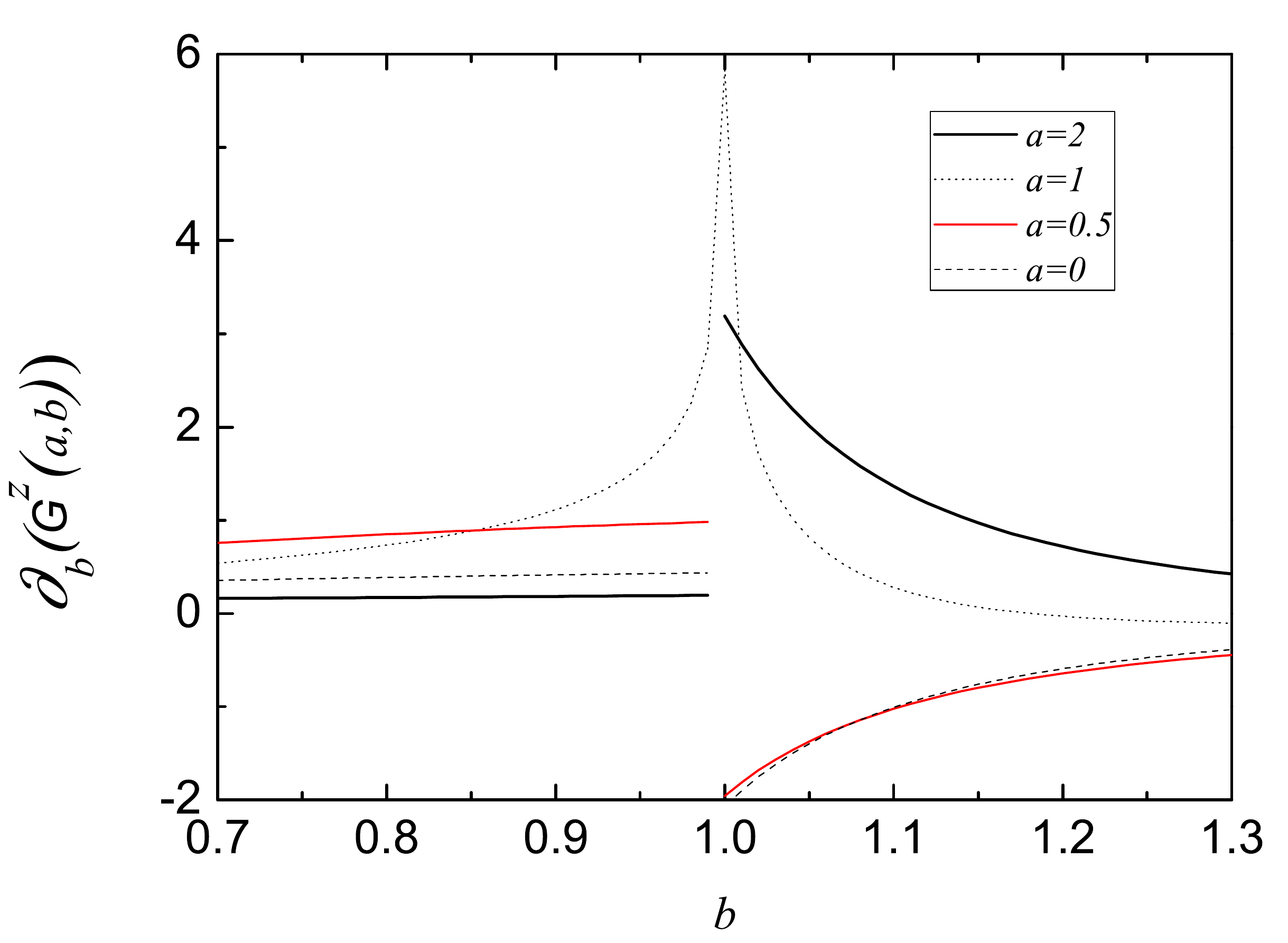}}
 \caption{$\gamma=0.6$. $\partial_{b} G^{z}_{1}(a,b)$ with fixed initial magnetic field $a=0, 0.5, 1,
2$.}\label{f10}
\end{figure}

In the following we use $\partial_{b} G^{z}_{1}(a,b)$ as an example
to analyze (see Fig.\ref{f10}). As we expect above, $\partial_{b}
G^{z}_{1}(a,b)$ is divergent at $b=1$ when $a=1$. While $a\neq1$, it
is discontinuous at $b=1$. How to explain the discontinuity is still
an open question to us.

\section{Summary}
In summary, starting from a general Hamiltonian which may undergo a
QPT with the change of a controllable parameter, we obtain a general
conclusion that in a sudden quench system, when the final
Hamiltonian is fixed, the behavior of the time-averaged expectation
of any observable has close relationship with the gapless excitation
of the initial Hamiltonian. This general conclusion, clarified by
our example of XY model, to a large extent explains the similarity
between the critical phenomena of sudden quench and static system.
The long-range correlation is also studied and we find that sudden
quench can destroy long-range correlation in the $x$ direction. As
we know, the one-spin reduced density matrix can always be written
as
$\rho=(I+\sum_{\alpha=x,y,z}\langle\sigma_{\alpha}\rangle\sigma_{\alpha})/2$.
If considering symmetry breaking effects in XY chain
\cite{PRA77032325}, we cannot suppose $\langle\sigma_{x}\rangle=0$.
Therefore for static case it will be a complex task to solve
$\langle\sigma_{x}\rangle$. But for sudden quench,
$\langle\sigma_{x}(a,
 b)\rangle=0$ for $a\neq b$. So
our result can help to simplify the calculation of one (two)-site
reduced density matrix. This will be useful in studying the
entanglement of one site with others for XY model. In addition, the
general conclusion can be useful for other quantities and other
models.

This work was supported by NSF of China under Grants No. 10821403
and No. 10974234, programs of Chinese Academy of Sciences, 973 grant
No. 2010CB922904 and National Program for Basic Research of MOST.

\newpage
\end{document}